\documentclass[onecollarge]{svjour2}       
\usepackage{graphicx,epsfig}

 \usepackage{amsfonts,amssymb}
\usepackage{amsmath}

\journalname{GRG}
\begin{document}

\title{Wave propagation in axion electrodynamics
} 


\author{Yakov Itin        
}


\institute{Institute of Mathematics, Hebrew University of Jerusalem, \at
            and Jerusalem College of Technology \\
              \email{itin@math.huji.ac.il}           
}

\date{Received: date / Accepted: date}

\maketitle

\begin{abstract}
In this paper, the axion contribution to the electromagnetic wave propagation is studied.
 First we show how the axion electrodynamics model can be  embedded into a premetric
formalism of Maxwell electrodynamics.
 In this formalism, the axion field is not an arbitrary added 
 Chern-Simon term of the Lagrangian, but emerges in a natural way as
 an irreducible part of a general constitutive tensor.
 We show that in order to represent the axion contribution to
 the wave propagation it is necessary
 to go beyond the geometric approximation, which is usually used in the premetric formalism.
 We derive a covariant dispersion relation for the axion modified
 electrodynamics.
 The wave propagation  in this model is studied for an
axion field with timelike, spacelike and null derivative covectors. The
birefringence effect emerges in all these classes  as a signal of
Lorentz  violation.
This effect is however completely different from the ordinary birefringence
appearing in  classical optics and in  premetric electrodynamics.
The axion field does not simple double the ordinary light cone structure. In fact, it modifies 
the global topological structure of light cones surfaces. In CFJ-electrodynamics, such a modification results in  violation of causality.
In addition, the optical metrics in axion electrodynamics are not pseudo-Riemannian. In fact, for all types of the axion field, they are even non-Finslerian.

\end{abstract}

\section{Introduction}
\label{intro} Axion electrodynamics, i.e., the standard electrodynamics
 modified by an additional axion
field \cite{Wilczek:1987mv},  is a subject of  a growing theoretical and
experimental interest. In particular, such a modification provides a
theoretical framework for a possible violation of parity and Lorentz invariance
 --- the Carroll-Field-Jackiw (CFJ) model \cite{Carroll:1989vb}, \cite{Jackiw:1998js}. For recent
developments of this model, see \cite{Kostelecky:2002hh} and the reference given therein.
Also the non-abelian extensions of the axion modified electrodynamics for the
Standard Model \cite{Colladay:1998fq} and gravity \cite{Kostelecky:2003fs}
 were worked out.

Although  Lorentz invariance is a basic assumption of the standard classical
and quantum field theory, in  quantum gravity and string theory this invariance
is probably violated. One believes that the low-energy manifestation of
Lorentz violation can be observed in experiments with the electromagnetic
waves. It justifies the importance of examination of the theoretical aspects of axion
contributions to the light propagation effects. Although this problem was
investigated intensively, we apply here an alternative approach based on a
metric-free (premetric) formulation of electrodynamics, see \cite{Birkbook}
 and the reference given therein.
 A characteristic feature of such a construction is
that the axion field is not involved by hand (merely as an additional
 term in the Lagrangian).
 Alternatively, in the premetric formalism, axion emerges as a
necessary and  natural partner of the standard photon variable.

In fact, there is a certain contradiction between the premetric
 electrodynamics and the CFJ-model. On one hand, in the premetric
 construction, one usually
states that the axion component does not alternate the wave propagation at all
\cite{Birkbook}. On the other hand, the Lorentz violation in the CFJ-model is
explicitly manifested in a modification of the standard light cone. In fact,
this contradiction comes from the specific geometric optics approximation which is
usually applied in the premetric electrodynamics.
 A constant axion field indeed
does not contribute to the wave propagation. To describe the wave propagation
in the CFJ-model one has to go beyond the geometric optics approximation and take
into account the first order derivatives of the axion field \cite{Itin:2004za},
\cite{serbia}.
With this modification, two constructions yield the equivalent results.

 The original CFJ-model is based on the standard special relativistic electrodynamics Lagrangian
 modified by an additional Chern-Simon term \cite{Carroll:1989vb}
\begin{equation}\label{intr1}
L=-\frac 14 F_{ij}F^{ij}-\frac 14p_i\epsilon^{ijkl}A_j\partial_kA_l+A_ij^i\,,
\end{equation}
where $p_i$ is a constant time-like vector. In fact, a special
 parametrization
$p_i=(\mu,0,0,0)$ is usually  used. Note that, in such a construction,
 the Lorentz
violation is involved by hand. The spatial $SO(3)$-invariance and gauge
invariance are preserved however. It was already pointed out in
\cite{Carroll:1989vb} that the model can be equivalently reformulated in an
explicitly gauge invariant form
\begin{equation}\label{intr2}
L=-\frac 14 F_{ij}F^{ij}+\psi\epsilon^{ijkl}F_{ij}F_{kl}+A_ij^i\,,
\end{equation}
where $\psi$ is a pseudoscalar axion field. It is related to the vector $p_i$
as
\begin{equation}\label{intr3}
\psi_{,i}=p_i\,,\qquad {\rm or} \qquad \psi=\mu t\,.
\end{equation}

In the current paper, we use a Lagrangian similar to (\ref{intr2}). We will not
restrict, however, to an axion field with   constant first order derivatives
and will not require it to be timelike. Moreover, we will consider a model on
a curved pseudo-Riemannian spacetime. Our approximation will be based,
however, on an assumption that the  gravitational field changes much more slowly
than the axion field. This restriction  considerably simplifies the
calculations and, hopefully, does not change the results, at least,
qualitatively.

The current paper is devoted to the 60th birthday  of Professor Bahram
Mashhoon. His permanent interest and considerable contribution to the study of
wave propagation effects are well known, see, for instance,
 \cite{Kopeikin:2001dz}, \cite{Hauck:2003gy}.
 The methods developed by
 Mashhoon will be useful also in the axion modified electrodynamics.
\section{The axion modified electrodynamics model}
\label{sec:1} 
Let us  briefly describe  how the axion modified
electrodynamics is embedded in the framework of the premetric
approach. We start with  two independent antisymmetric tensor fields,
 the field strength tensor $F_{ij}$ and the excitation field ${\mathcal H}^{ij}$. The latter  field is a pseudotensor density of weight $(+1)$.
The flux conservation law (the first Maxwell equation) is postulated,
\begin{equation}\label{field-eq0}
\epsilon^{ijkl}F_{ij,k}=0\,.
\end{equation}
 Here  the Roman indices  range from 0 to 3, $\epsilon^{ijkl}$
and $\epsilon_{ijkl}$ are the Levi-Civita permutation tensors normalized
with
  $\epsilon^{0123}=-\epsilon_{0123}=1$, the commas denote the ordinary partial 
  derivatives.
Due to (\ref{field-eq0}), the field strength tensor is expressed in term of
the standard vector potentials $A_i$,
\begin{equation}\label{potential}
F_{ij}=\frac 12 \left(A_{i,j}-A_{j,i}\right)\,.
\end{equation}

A local linear homogeneous constitutive relation between the fields  $F_{ij}$
and  ${\mathcal H}^{ij}$,
  \begin{equation}\label{rel}
{\mathcal H}^{ij}=\frac 12 \chi^{ijkl}F_{kl}\,,
\end{equation}
is assumed. By  definition, the constitutive pseudotensor $\chi^{ijkl}$ must  respect the symmetries of the fields $F_{ij}$  and ${\mathcal H}^{ij}$,
\begin{equation}\label{sym}
\chi^{ijkl}=\chi^{[ij]kl}=\chi^{ij[kl]}\,.
\end{equation}
Hence  it   has, in general,  36 independent components. The irreducible
decomposition of this tensor under the group of linear transformations
involves three independent pieces. One of these three pieces is the axion field, which is a
subject of our interest.

The high number of components of $\chi^{ijkl}$ allows to describe a wide range
of physical effects. The axion field, however, adds only one addition component
to the standard electrodynamics. So, in order to embed the axion electrodynamics into the premetric approach, one has to restrict the number of the independent components.  The first restriction  comes from the Lagrangian formulation of the
model. We assume the action integral to be of the standard form
\begin{equation}\label{lagr1}
{\mathcal A}=\int_M \left(F_{ij}{\mathcal H}^{ij} + A_i{\mathcal J}^{i}\right)d^4x\,.
\end{equation}
When (\ref{rel}) is substituted, the action takes the form
\begin{equation}\label{lagr2}
{\mathcal A}=\int_M \left(\frac 12 \chi^{ijkl}F_{ij}F_{kl}+ A_i{\mathcal J}^i\right)d^4x\,.
\end{equation}

Note, however, that, in contrast to the ordinary textbooks formulation,  the
electromagnetic current ${\mathcal J}^i$ and the excitation field ${\mathcal H}^{ij}$ are  pseudotensorial densities of  weight $+1$ (see Appendix for definition and  details).
 Thus the Lagrangian (\ref{lagr1}) is general
covariant and admits arbitrary transformations of coordinates. The constitutive
pseudotensor involved in (\ref{lagr1}) respects an additional symmetry
\begin{equation}\label{add-sym}
\chi^{ijkl}=\chi^{klij}\,.
\end{equation}
This condition removes 15 independent components of $\chi^{ijkl}$ which compose
the so-called skewon part of the constitutive tensor. The remaining quantity of
21 independent components is decomposed irreducibly to a principle part of 20
components plus one component that represents the axion field. In contrast to
the Lagrangian (\ref{intr2}), the premetric formulation (\ref{lagr1}) does not
involve an addition axion term in the Lagrangian. In fact, such a term  is hidden in the
constitutive tensor.

The variation of (\ref{lagr1}) with respect to the vector potentials $A_i$
yields the second Maxwell equation
\begin{equation}\label{field-eq}
 {\mathcal H}^{ij}_{}{,j}={\mathcal J}^i\,, \qquad {\rm or}\qquad
 \frac 12 \left(\chi^{ijkl}F_{kl}\right)_{,j}={\mathcal J}^i\,.
\end{equation}

  Observe that this general construction, is explicitly gauge invariant.
  As usually, the  charge conservation law $\partial_i {\mathcal J}^i=0$ is a
  consequence of the  field equation (\ref{field-eq}).

 The standard  electrodynamics is reinstated in this
formalism  by a special Maxwell constitutive tensor
\begin{equation}\label{Max-const-tensor}
^{\tt (Max)}\chi^{ijkl}=\left(g^{ik}g^{jl}-g^{il}g^{jk}\right)\sqrt{-g}\,.
\end{equation}
Here $g^{ij}$ is a metric tensor with the signature $\{+,-,-,-\}$ and with the
determinant $g$. 
To involve the axion field
contribution to the standard electromagnetic Lagrangian, it is enough now to
consider a slightly modified constitutive tensor of the following form
\begin{equation}\label{const-tensor}
\chi^{ijkl} ={}^{\tt (Max)}\chi^{ijkl}+ \psi\epsilon^{ijkl}\,.
\end{equation}
Here  the axion $\psi$ is a  pseudoscalar field. It is invariant under transformations of coordinates with positive determinant and changes its sign under transformations with negative determinant. 

The constitutive tensor  (\ref{const-tensor}) is not merely a modification of
the standard expression (\ref{Max-const-tensor}).
 In fact, it is an irreducible decomposition of a
general constitutive tensor $\chi^{ijkl}$.
 The Lagrangian formulation removes the skewon part
of it. An additional requirement of closeness, see \cite{Birkbook},
restricts the principle part of 20 independent components  to a pure metrical expression. So the axion part
appears in this formalism as a natural and necessary ingredient of a general
construction. 
To simplify our consideration we will treat the axion field only
phenomenologically. An additional dynamical axion Lagrangian can be readily
added to (\ref{lagr1}).

\section{Wave solution}\label{sec:2}

To study the wave propagation in the axion modified model, we assume the
electromagnetic current to be equal to zero. Substitute (\ref{potential}) into
the second field equation (\ref{field-eq}) to rewrite it as
\begin{equation}\label{fe-2}
\left(\chi^{ijkl}A_{k,l}\right)_{,j}=0\,,
\end{equation}
or, equivalently, as 
\begin{equation}\label{fe-2x}
\chi^{ijkl}A_{k,lj}+\chi^{ijkl}{}_{,j}A_{k,l}=0\,.
\end{equation}
In this paper, we will apply the following approximation
\begin{equation}\label{approx}
\chi^{ijkl}{}_{,j}=\psi_{,j}\epsilon^{ijkl}\,.
\end{equation}
In other words, we restrict to the spacetime regions  where the
gravitational field varies slowly as compared to the change of the pseudoscalar
field.  In particular, our analysis will be exact for an axion field
considered on a flat Minkowski space.

Substituting (\ref{const-tensor}) and (\ref{approx}) into (\ref{fe-2x}) we get
\begin{equation}\label{fe-2xx}
\left(g^{ik}g^{jl}-g^{il}g^{jk}\right)A_{k,lj}\sqrt{-g}+
\psi_{,j}\epsilon^{ijkl}A_{k,l}=0\,.
\end{equation}
 We are looking for a plane monochromatic  wave solution of the equation
(\ref{fe-2xx}). Write it as
\begin{equation}\label{sol}
A_k={\rm Re}\left(a_ke^{iq_mx^m}\right)\,.
\end{equation}
Here the amplitude $a_k$ and the wave covector $q_m$ do not depend on a point.
Both quantities can be  complex, the physical solution $A_k$ is equal
 to the real
part of the corresponding complex expression. Since (\ref{fe-2xx}) is a linear field equation
 with real coefficients, it is possible to deal, as usual,
 with the complex valued expression  $A_k=a_ke^{iq_mx^m}$ itself. 
Substituting this ansatz in
(\ref{fe-2xx}) we have
\begin{equation}\label{fe-3}
\left(g^{ik}g^{jl}-g^{il}g^{jk}\right)\sqrt{-g}q_jq_la_k-i\psi_{,j}q_l
\epsilon^{ijkl}a_k=0\,.
\end{equation}
This tensorial equation is represented by  a linear system of four equation for
four components of the covector $a_k$. Write it briefly as
\begin{equation}\label{fe-3-new}
M^{ij}a_j=0\,,
\end{equation}
where
\begin{eqnarray}\label{M-matrix}
M^{ij}&=&\left(g^{ij}g^{kl}-g^{il}g^{jk}\right)\sqrt{-g}q_kq_l+
i\psi_{,k}q_l\epsilon^{ijkl}\nonumber\\
&=&\left(g^{ij}q^2-q^iq^j\right)\sqrt{-g}+i\psi_{,k}q_l\epsilon^{ijkl}\,.
\end{eqnarray}
Observe two  evident relations that hold due to the definition of the matrix
$M^{ik}$
\begin{equation}\label{M-relations}
M^{ij}q_i=0\,,\qquad M^{ij}q_j=0\,.
\end{equation}
These relations can be given a pure physical sense: The former one represents
the charge conservation law, while the latter relation represents  the gauge
invariance of the field equation.

  The linear system
(\ref{fe-3-new}) has a non-zero solution if and only if its determinant equal to
zero.  For the system (\ref{fe-3-new}), this condition holds identically, which can
be seen even without explicit calculation of the determinant.
 Indeed, the identities
(\ref{M-relations}) express  linear relation between the rows (and between the
columns) of the matrix $M^{ik}$. So this matrix is singular. 
Moreover, (\ref{M-relations}$_b$) also means  that the linear system
(\ref{fe-3-new}) has a non-zero solution of the form
\begin{equation}\label{trivial-solut}
a_j=Cq_j\,
\end{equation}
with an arbitrary constant $C$. This solution is evidently unphysical  since
it corresponds to the gauge invariance of the field equations. To describe an
observable  physically meaningful situation we must have  an additional linear
independent solution of (\ref{fe-3-new}).
\section{Dispersion relation}
The linear system (\ref{fe-3}) has two linear independent solutions (one for
gauge and one for physics) if and only if its matrix $M^{ij}$ is of rank 2 or
less. An algebraic expression of this requirement is
\begin{equation}\label{Adj1}
A_{ij}=0\,,
\end{equation}
 where $A_{ij}$ is the adjoint matrix.
This matrix is obtained by removing the $i$-th row and the $i$-th column
from the original matrix. The determinants of the retaining $3\times 3$
matrices are calculated and assembled in a new matrix $A_{ij}$.
 The entries of
the adjoint matrix are expressed via the entries of the matrix $M^{ij}$ as
\begin{equation}\label{Adj2}
A_{ij}=\frac 1{3!}
\epsilon_{ii_1i_2i_3}\epsilon_{jj_1j_2j_3}M^{i_1j_1}M^{i_2j_2}M^{i_3j_3}\,.
\end{equation}
Note that for a covariant tensor of a rank $(2,0)$, the adjoint matrix is a contravariant tensor of a rank $(0,2)$. 
 Since the adjoint matrix has,
in general, 16 independent components it seems that we have to require 16
independent conditions. The following algebraic fact \cite{Itin:2007av} shows that the situation is rather simpler.

{\bf  Proposition:} {\it If a square $n\times n$ matrix $M^{ij}$
  satisfies the relations
   \begin{equation}\label{matr-eq}
M^{ij}q_i=0\,,\qquad M^{ij}q_j=0\,
  \end{equation}
  for some nonzero vector $q_i$,
  its adjoint matrix $A_{ij}$ is represented by}
   \begin{equation}\label{Adj2-new}
A_{ij}=\lambda(q)q_iq_j\,.
   \end{equation}
   For a formal proof of this fact, see \cite{itin}.
Consequently, instead of 16 conditions (\ref{Adj1}),  we have to require only one condition
\begin{equation}\label{gen-dis}
\lambda(q)=0\,.
\end{equation}
Recall that this condition is necessary for existence a physically meaningful
solution of the generalized wave equation, so it is  a generalized dispersion
relation.

We calculate now the adjoint matrix for the axion modified electrodynamics
model. Substituting (\ref{M-matrix}) into (\ref{Adj2}) we get
\begin{eqnarray}\label{Adj3}
A_{ij}&=&\frac 1{3!} \epsilon_{ii_1i_2i_3}\epsilon_{jj_1j_2j_3}
\left({\cal M}^{i_1j_1}+i\psi_{,k_1}q_{l_1}\epsilon^{i_1j_1k_1l_1}\right)\nonumber\\
&&\times\left({\cal
M}^{i_2j_2}+i\psi_{,k_2}q_{l_2}\epsilon^{i_2j_2k_2l_2}\right) \left({\cal
M}^{i_3j_3}+i\psi_{,k_3}q_{l_3}\epsilon^{i_3j_3k_3l_3}\right)\,,
\end{eqnarray}
where  a notation
\begin{equation}\label{MAX-M}
{\cal M}^{ij}=\left(g^{ij}q^2-q^iq^j\right)\sqrt{-g}\,
\end{equation}
for the pure Maxwell part of the matrix $M^{ij}$ is involved.

Calculate now in turn the entries of (\ref{Adj3}) as the powers of the
imaginary unit. The calculations are considerable simplified when we take into
account that due to the Proposition above the result has to be a symmetric
matrix. It means that all  antisymmetric contributions to $A_{ij}$ are
canceled. 
The   free in $i$ term, i.e. the pure Maxwell term, takes the form
\begin{equation}\label{Adj4}
\frac 1{3!} \epsilon_{ii_1i_2i_3}\epsilon_{jj_1j_2j_3}{\cal M}^{i_1j_1}{\cal
M}^{i_2j_2}{\cal M}^{i_3j_3}=-\sqrt{-g}q^4q_iq_j\,.
\end{equation}
The $i$-term takes the form
\begin{equation}\label{Adj5}
i(1/2)\epsilon_{ii_1i_2i_3}\epsilon_{jj_1j_2j_3}{\cal M}^{i_1j_1}{\cal
M}^{i_2j_2}\psi_{,k_3}q_{l_3}\epsilon^{i_3j_3k_3l_3}\,.
\end{equation}
Since the Maxwell matrix ${\cal M}^{ij}$ is symmetric, this expression is
antisymmetric in the indices $i,j$ and does not give a contribution to the
adjoint matrix. In fact, explicit calculations show that the expression
(\ref{Adj5}) vanishes identically. 
 The $i^2$-term is given by 
\begin{eqnarray}\label{Adj6}
&&i^2\,(1/2)\epsilon_{ii_1i_2i_3}\epsilon_{jj_1j_2j_3}{\cal
M}^{i_1j_1}\psi_{,k_2}q_{l_2}\epsilon^{i_2j_2k_2l_2}\psi_{,k_3}q_{l_3}
\epsilon^{i_3j_3k_3l_3}\nonumber\\
&&=-\sqrt{-g}\left[\left(\psi_{,m}\psi^{,m}\right)q^2-
\left(\psi_{,m}q^m\right)^2\right]q_iq_j\,.
\end{eqnarray}
The $i^3$-term takes the form
\begin{equation}\label{i3}
(i)^3(1/2)\epsilon_{ii_1i_2i_3}\epsilon_{jj_1j_2j_3}\psi_{,k_1}\psi_{,k_2}\psi_{,k_3}q_{l_1}q_{l_2}q_{l_3}
\epsilon^{i_1j_1k_1l_1}\epsilon^{i_2j_2k_2l_2}\epsilon^{i_3j_3k_3l_3}
\end{equation}
This expression is evidently antisymmetric in the indices $i,j$ so it does not
give a contribution to the symmetric matrix $A_{ij}$. In fact, it is equal to
zero.

Consequently we have the adjoint matrix  in the following form
\begin{equation}\label{lambda1}
A_{ij}=-\sqrt{-g}\left[q^4+\left(\psi_{,m}\psi^{,m}\right)q^2-
\left(\psi_{,m}q^m\right)^2\right]q_iq_j\,.
\end{equation}
Thus the dispersion relation for the electromagnetic waves in the axion electrodynamics is expressed as
\begin{equation}\label{disp}
q^4+\left(\psi_{,m}\psi^{,m}\right)q^2-\left(\psi_{,m}q^m\right)^2=0\,.
\end{equation}
 Observe some characteristic features of this equation.

{\bf{1.}}  In Minkowski space, for an axion field with a constant covector of derivatives, it coincides with the dispersion relation expression given in \cite{Carroll:1989vb}.

{\bf{2.}} The axion dispersion relation (\ref{disp}) is essentially different from the general covariant
 dispersion relation appearing in the premetric electrodynamics.
 The premetric  dispersion relation is a quartic homogeneous polynomial in the wave covector variable $q$. Its general form is
\begin{equation}\label{disp-prem}
{\cal G}^{ijkl}q_iq_jq_kq_l=0\,.
 \end{equation}
 Certainly the homogeneity is originated in  the geometric approximation used for its derivation. The quartic polynomial  of the axion electrodynamics (\ref{disp}) is not homogeneous, so it provides some additional types of light cones structure. In particular, the birefringence effect here comes from the derivatives of the media parameters.  On the contrary, in the premetric electrodynamics as well as in the classical crystal optics, the birefringence effect comes from the tensorial nature of the parameters  of the media.

{\bf{3.}} Another interesting feature of the dispersion relation (\ref{disp}) is that it is, in fact, a general covariant expression. Indeed it is invariant under general pointwise transformations of coordinates (with positive or negative determinant).  This is in spite of the fact that we used for its derivation  the ansatz (\ref{sol}) which is only
 special relativistic.

{\bf 4.} The relation (\ref{disp}) is invariant under the transformation
 $q^i\to -q^i$.  Thus the light cones structure has to be $PT$ invariant.

 {\bf 5.} The relation (\ref{disp}) is invariant under the transformation
 of the field $\psi^i\to -\psi^i$. Thus the light cones have to be similarly
 oriented relative to the vector $\psi^i$.

 {\bf 6.} A quartic expression dispersion relation (\ref{disp}) is  sometimes  factored to a product of two second order polynomials. It is easy to see that the relation (\ref{disp}) cannot be factored in covariant way. Such factorization is possible, however, in special coordinates. Since (\ref{disp}) is not homogeneous, at least one of these quadratic factors is not homogeneous.

 {\bf 7.} For ordinary electrodynamics in vacuum, the factors are homogeneous and coincide. This unique homogeneous quadratic factor corresponds to unique  Minkowski metric. For  general anisotropic media, in  (geometric approximation) two  factors are homogeneous but different one from another. This case is formulated in term of two optical metrics, both pseudo-Riemannian. In axion electrodynamics, at least one factor is necessary inhomogeneous. This factor cannot be reformulated in term of on ordinary pseudo-Riemannian metric. In fact, it does not even correspond to a general Finslerian metric, which in general appear in premetric electrodynamics \cite{Perlick},\cite{Perlick:2005hz}.
\section{Special axion fields}
\subsection{Axion field on Minkowski spacetime}

In this section, we restrict for simplicity to the Minkowski spacetime.
 In the  Cartesian coordinates, $g^{ij}={\rm diag}(1,-1,-1,-1)$ with $g=-1$, so the
dispersion relation (\ref{disp}) takes the form
\begin{equation}\label{disp-M}
q^4+\left(\psi_{,m}\psi^{,m}\right)q^2-\left(\psi_{,m}q^m\right)^2=0\,.
\end{equation}
Note that now all scalar products are taken with respect to the constant Minkowski metric.
We  apply the $(1+3)$ splitting and  denote
\begin{equation}\label{q-spli}
q=(w,{\mathbf{k}})\,,\qquad k=|{\mathbf{k}}|\,,
\end{equation}
and
\begin{equation}\label{psi-spli}
\psi_{,i}=(\mu,{\mathbf{m}})\,,\qquad m=|{\mathbf{m}}|\,.
\end{equation}
Denote  by $\alpha$ the angle between the vectors $\mathbf{m}$ and $\mathbf{k}$. For a complex vector $q^i$, the usual analytic extension for $\alpha$ is assumed.
 Due to the symmetry of (\ref{disp-M}) under the reflection
$\psi_i\to -\psi_i$, we can deal locally with the case $\mu>0$.

In the $(1+3)$ notation, the dispersion relation (\ref{disp-M}) takes the form
\begin{equation}\label{disp-M-split}
(w^2-k^2)^2+ \left(\mu^2-m^2\right)(w^2-k^2)-
\left(w\mu+mk\cos\alpha\right)^2=0\,.
\end{equation}
 It is useful to express the dispersion relation in term of the phase velocity $v_p=w/k$
\begin{equation}\label{disp-M-x}
v_p^4-v_p^2\left(2+\frac{m^2}{k^2}\right)-2v_p\frac{\mu m}{k^2}\cos\alpha
+\left(1-\frac{\mu^2}{k^2}+\frac{m^2}{k^2}\sin^2\alpha\right)=0\,.
\end{equation}
 We observe now that, in the left hand side of (\ref{disp-M-x}), the dependence on the angle $\alpha$ can be removed only if the vector $ \mathbf{m}$ can be taken equal to zero.
 Such coordinates can be chosen  if and only if the 4-covector $\psi_{,i}$ is timelike.
 In this case, the $SO(3)$ invariance is preserved.
 Alternatively, for an arbitrary  null or spacelike covector $\psi_{,i}$,
 the rotational symmetry is violated.

 Additionally, let the term linear  in $v_p$  cannot be removed, i.e.,  the parameters $m$ or $\mu$ cannot be chosen be equal to zero. Consequently,  $v_p$ and  $(-v_p)$ cannot  satisfy simultaneously the same dispersion relation.
 It means that the future and the past light cones are not identical.
 Thus the time inversion symmetry ($T$-invariance) is violated.
 Since the whole equation (\ref{disp-M}) is $PT$-invariant,
 the parity invariance ($P$-invariance) is also violated.

 The axion field and the
covector of its derivatives are assumed to change smoothly in the whole spacetime.
Thus the spacetime itself is separated to distinct regions with different norms of the
covector $\psi_m$. In every specific region,  special coordinates can be chosen in
order to simplify the parametrization of the covector field $\psi_m$.

\subsection{Axion field with a timelike derivative}
This model and its physics consequences   was studied in the original version of axion modified
 electrodynamics
- the Carroll-Field-Jackiw model \cite{Carroll:1989vb}. Consider a spacetime
region, where the derivatives of the axion field compose a timelike covector
\begin{equation}\label{psi0x}
\psi_{,i}\psi^{,i}>0\,.
\end{equation}
Choose in this region the time coordinate axis to be directed
 along the covector $\psi_{,i}$.
 Consequently, this covector is parameterized now as
\begin{equation}\label{psi0}
\psi_{,i}=(\mu,0,0,0)\,.
\end{equation}
 Due to the symmetries of the dispersion relation (\ref{disp}), we can,
 without lose of generality, require $\mu>0$.
 In the original CFJ-model, the axion field
was given as $\psi=\mu t$ with a constant parameter $\mu$. Observe, however,
 that only the
first order derivatives of the axion field are involved in the dispersion
relation (\ref{disp}). So, in fact, we can deal with a more general case, where
$\mu$ is an arbitrary function of a point.

Substituting (\ref{psi0}) into
(\ref{disp-M-split}) we get
\begin{equation}\label{psi0-0}
 (w^2-k^2- \mu k)( w^2-k^2+ \mu k)=0\,.
 \end{equation}
The solutions of this equation are
\begin{eqnarray}\label{four-sol-1}
{}^{(1)}w&=&\sqrt{k^2+\mu k}\,,\qquad {}^{(2)}w=-\sqrt{k^2+\mu k}\,,\\
\label{four-sol-2}
 {}^{(3)}w&=&\sqrt{k^2-\mu k}\,,\qquad
{}^{(4)}w=-\sqrt{k^2-\mu k}\,.
 \end{eqnarray}
  Thus, for $k>\mu$, 
 we have four distinct  real solutions: two positive and two negative.
 For $k=\mu$, there are  two real solutions of opposite signs and one double solution equal to zero. For $k<\mu$, two solutions are real and two are pure imaginary.

 Geometrically, the  solutions (\ref{four-sol-1}, \ref{four-sol-2}) define  two distinct double hypersurfaces
 \begin{equation}\label{hyper1}
w^2-k_1^2-k_2^2-k_3^2- \mu\sqrt{k_1^2+k_2^2+k_3^2}=0\,.
\end{equation}
and
 \begin{equation}\label{hyper1x}
w^2-k_1^2-k_2^2-k_3^2+ \mu\sqrt{k_1^2+k_2^2+k_3^2}=0\,.
\end{equation}
\begin{figure*}[h]
$\!\!\!\!$\begin{tabular}{ccc}
\begin{minipage}{3.0in}
\centering
\includegraphics[angle=0,width=3.0in]{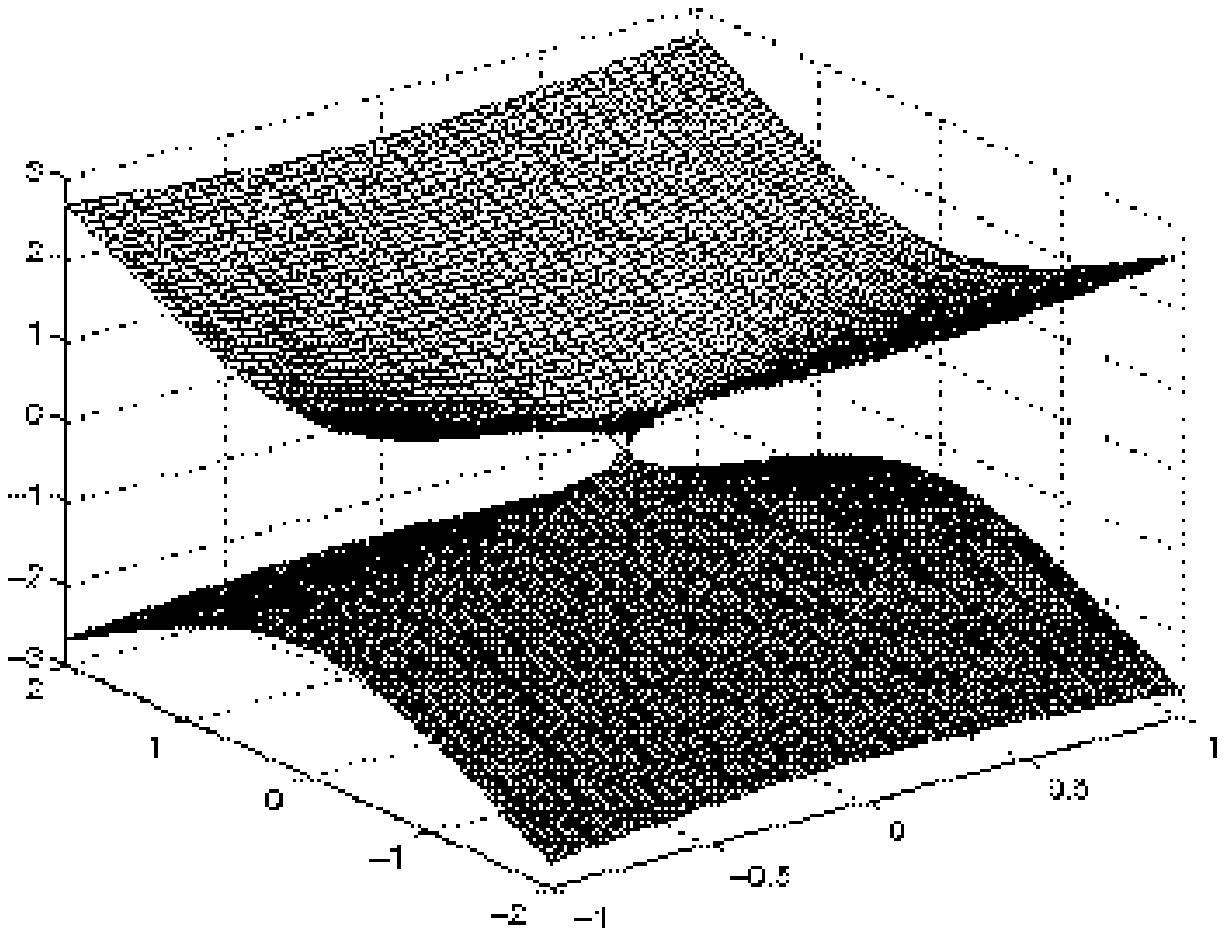}
\end{minipage}
&
\begin{minipage}{3.0in}
\centering
\includegraphics[angle=0,width=3.0in]{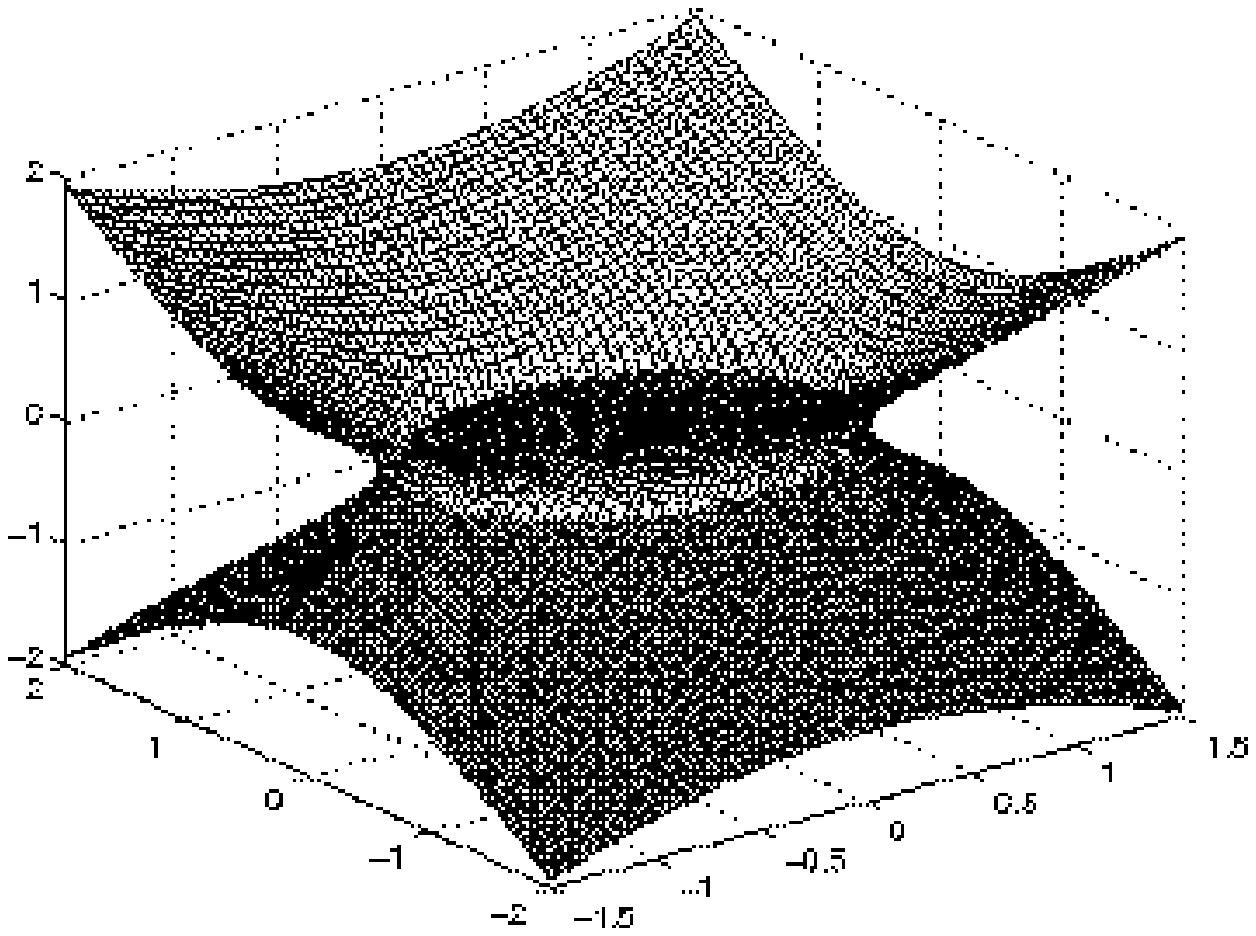}
\end{minipage}
\end{tabular}
\caption{ The graphs represent two hypersurfaces corresponding to CFJ modified
electrodynamics}  \label{cone1}
\end{figure*}

We depict on Fig. 1 these hypersurfaces in the coordinates $(k_1, k_2, w)$ (the third coordinate $k_3$ is suppressed, as usual).
The first picture corresponds to the solution (\ref{four-sol-1}). It is topologically equivalent to the ordinary light cone. In particular, two cones are joined by a unique point $(w=0, {\mathbf{k}}=0)$. When this point is removed the interior region bounded by the hypersurface is separated to two disjoint parts   -- the future and the past spacetime regions. The second hypersurface corresponding to (\ref{four-sol-2}) is topologically different from the ordinary light cone structure. It consists of two pieces which are joined by  a sphere  $(k=\mu , w=0)$ and by a point $(k=0 , w=0)$.
Consequently the interior region cannot be separated into two disjoint parts even if the origin is removed. It means that the future (the upper region) and the past (the downer region) always connected and cannot be disjoint. In other words, the causality on this branch is violated.

 Since the equations  (\ref{hyper1},\ref{hyper1x}) involve a term linear in the covector $k$ the corresponding  light hypersurfaces cannot be associated with some pseudo-Riemannian  optical metrics.

 The  phase velocities of the waves are given by the expressions
\begin{equation}\label{psi0-1}
{}^{(1)}v_p=\sqrt{1+\frac{\mu}k}\,,\qquad {}^{(2)}v_p=\sqrt{1-\frac{\mu}k}\,,
\end{equation}
which coincide with the corresponding  formulas of \cite{Carroll:1989vb}.
  One of the phase velocities is greater than the speed of light in a vacuum.
  It increases monotonically when the factor $\mu/k$ increases, i.e., when the parameter $k$ tends to zero.
 The second phase velocity is less than the speed of light in a vacuum and monotonically decreases  to zero when the parameter $k$ tends  to zero.

 Well known that the phase velocity  does not completely characterize
 the energy propagation.
 Another   useful characteristic  is
the group velocity which is usually  thought of as the velocity
 at which energy is propagated along a wave. This quantity is defined  by the derivative
\begin{equation}\label{group-velo}
v_g=\frac {\partial w}{\partial k}\,.
\end{equation}
From (\ref{psi0-1}) we have
\begin{equation}\label{group-velo-x}
{}^{(1)}v_g=\frac {k+\mu/2}{\sqrt{k^2+ \mu k}}\,,\qquad
{}^{(2)}v_g=\frac {k-\mu/2}{\sqrt{k^2- \mu k}}\,.
\end{equation}
Consequently, both group velocities are superluminal and monotonically increase when the parameter $k$ tends to zero.  Observe that on the second branch, both velocities are defined only for $0\le \mu/k\le 1$. For $k\to \mu$, the phase velocity tends to zero while the group velocity goes to infinity. This behavior corresponds to transmission of energy with infinite velocity, i.e., indicates the runaway modes, see \cite{Carroll:1989vb} and \cite{Adam:2001ma}.

Consequently, the waves with $k>\mu$  propagate along two distinct future light hypersurfaces.
This behavior corresponds to the birefringence effect known from classical
optics. The topological type of the light hypersurfaces is different, however, from the ordinary light cones. For $k<\mu$,  the runaway modes emerge.

\subsection{Axion field with a spacelike derivative}
Let us consider a  spacetime region where
 the axion field has a spacelike covector of
derivative
\begin{equation}\label{psi1x}
\psi_{,m}\psi^{,m}<0\,.
\end{equation}
 By transformation of the coordinates, we can choose in the whole region a
parametrization
\begin{equation}\label{psi1}
\psi_{,i}=(0,m,0,0)\,.
\end{equation}
Also here the parameter $m$ can be considered as  a function of a point. We can restrict to  $m>0$.
 The dispersion relation (\ref{disp-M-split})
takes now the form
 \begin{equation}\label{disp-spacelike}
 w^2=k^2+\frac {m^2}2\pm\sqrt{\frac{m^4}4+m^2k^2\cos^2\alpha}\,.
 \end{equation}
 Observe that an inequality $w^2\ge k^2\ge 0$ holds for (\ref{disp-spacelike}).
 Hence  this equation
 for $w$ has four real  solutions  for every values of parameters.
 Consequently the runaway solutions are absent.

The corresponding light hypersurfaces are given by
\begin{equation}\label{hyper-spacelike1}
w^2-k_1^2-k_2^2-k_3^2-\frac {m^2}2\left(1+
\sqrt{1+\frac{4k_1^2}{m^2}}\right)=0\,.
\end{equation}
and
\begin{equation}\label{hyper-spacelike2}
w^2-k_1^2-k_2^2-k_3^2-\frac {m^2}2\left(1-
\sqrt{1+\frac{4k_1^2}{m^2}}\right)=0\,.
\end{equation}
\begin{figure*}[h!]

$\!\!\!\!$\begin{tabular}{ccc}
\begin{minipage}{3.0in}
\centering
\includegraphics[angle=0,width=3.0in,scale=0.5]{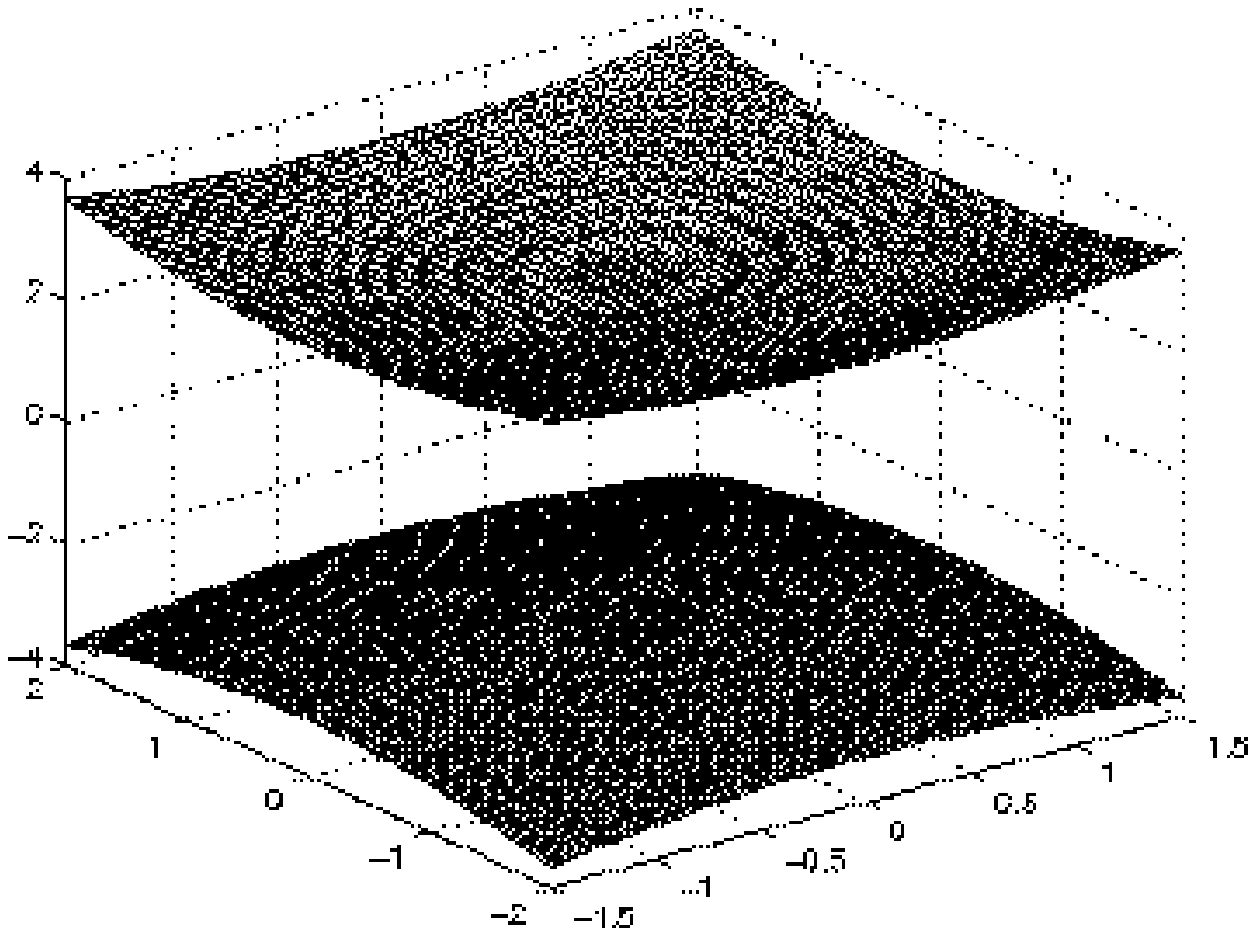}
\end{minipage}
&
\begin{minipage}{3.0in}
\centering
\includegraphics[angle=0,width=3.0in,scale=0.5]{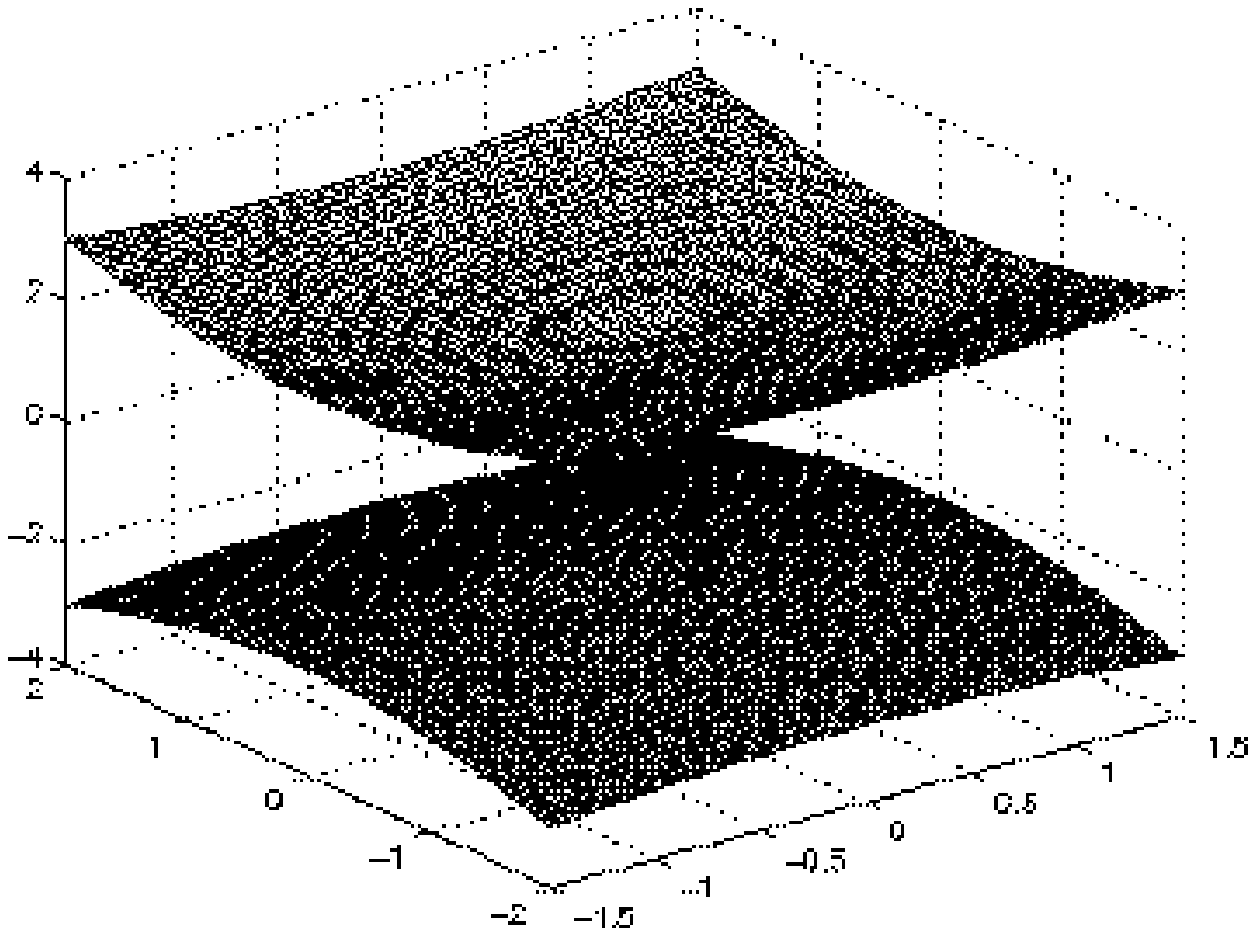}
\end{minipage}
\end{tabular}
\caption{The graphs represent two double light hypersurfaces of the spacelike axion
modified electrodynamics. } \label{cone2-new}
\end{figure*}

These hypersurfaces are depicted on Fig.2.  The first structure is topologically different from the standard one. Indeed, (\ref{hyper-spacelike1}) has not real solutions for $w=0$. Consequently, two branch given by (\ref{hyper-spacelike1}) are represented by two disjoint surfaces. In other words there is not, for this lightlike hypersurfaces, a way from the past to the future.  The second equation (\ref{hyper-spacelike2}), for $w=0$, has a unique solution $k_1=k_2=k_3=0$.  Thus we have on this branch the standard light cone topology with one point joining the past and the future spacetime regions. 
However, an ordinary definition of the causality is not applicable in this case. Indeed, due to the inhomogeneity of the dispesion relation,  a  type of a light trajectory depends on the parametrization.  
The expressions  (\ref{hyper-spacelike1}, \ref{hyper-spacelike2}) contain non-polynomial terms. Thus also in this case, the optical metrics cannot be represented in a  pseudo-Riemannian form. 

In term of the phase velocity, (\ref{disp-M-x}) is rewritten as
\begin{equation}\label{disp-M-xx}
v_p^4-v_p^2\left(2+\frac{m^2}{k^2}\right)
+\left(1+\frac{m^2}{k^2}\sin^2\alpha\right)=0\,.
\end{equation}
 Thus the phase velocities are expressed as
\begin{equation}\label{sol-M-xx}
{}^{(1)}v_p=\sqrt{1+\frac{m^2}{2k^2}+\sqrt{\frac{m^2}{k^2}\left(\frac{m^2}{4k^2}+
\cos^2\alpha\right)}}\,
\end{equation}
and
\begin{equation}\label{sol-M-xxx}
{}^{(2)}v_p=\sqrt{1+\frac{m^2}{2k^2}-\sqrt{\frac{m^2}{k^2}\left(\frac{m^2}{4k^2}+
\cos^2\alpha\right)}}\,.
\end{equation}
These expressions   depend explicitly  on the angle $\alpha$, so the $SO(3)$ invariance is violated. However, due to the fact that  the expression is invariant under the change
 $v_p\to -v_p$, the future and the past cone are the same. Thus the
 $T$-invariance and consequently the $P$-invariance are preserved.
For small $k$ the first phase velocity (\ref{sol-M-xx}) goes to infinity while the second phase velocity (\ref{sol-M-xxx}) tends to zero.
In the  transversal direction, $\alpha=\pi/2$,
 one of the phase velocities is greater
 and one is equal to the speed of light in a vacuum.

 The group velocities are expressed as
\begin{equation}\label{group}
v_g=\frac 1{v_p}\left(1\pm\frac {2\cos^2\alpha}{\sqrt{1+\frac{4k^2}{m^2}\cos^2\alpha}}\right)\,.
\end{equation}
Both expressions tend to zero for small values of $k$.
In transversal direction we have
 \begin{equation}\label{grou1}
 v_g=\frac 1{v_p}\,,
 \end{equation}
 i.e.,  one of the group velocities is less
 and one is equal to the speed of light in vacuum.

\subsection{Axion field with a lightlike derivative}
Consider a spacetime region, where the derivatives of the axion field
 compose a null covector
\begin{equation}\label{psi1x-new}
\psi_{,m}\psi^{,m}=0\,.
\end{equation}
In this case,  the general covariant dispersion relation (\ref{disp})
 takes the form
\begin{equation}\label{disp-nul}
q^4-\left(\psi_{,m}q^m\right)^2=0\,.
\end{equation}
 This expression is readily factored as
 \begin{equation}\label{disp-nul1}
(q^2-\psi_{,m}q^m)(q^2+\psi_{,m}q^m)=0\,.
\end{equation}
The expression in the left hand side does not have a defined sign
 for an arbitrary non zero covector $q^m$.
 Thus the non-birefringence condition \cite{Itin:2005iv} is
 explicitly violated. Consequently, the birefringence effect
 emerges for arbitrary varying null axion fields.

 We can choose a parametrization
\begin{equation}\label{psi1-new}
\psi_{,i}=(m,m,0,0)\qquad m>0\,.
\end{equation}
 On a flat Minkowski manifold, the dispersion relation takes now the form
  \begin{equation}\label{disp-nul2}
(w^2-k^2)^2-m^2\left(w+k\cos\alpha\right)^2=0\,,
\end{equation}
 or,
  \begin{equation}\label{disp-nul3}
w^2=k^2 \pm m(w+k\cos\alpha)\,,
\end{equation}
Consequently the light hypersurfaces are expressed as
\begin{equation}\label{hyper-nul3}
w^2-k_1^2-k_2^2-k_3^2 + m(w+k_1)=0\,,
\end{equation}
and
\begin{equation}\label{hyper-nul3x}
w^2-k_1^2-k_2^2-k_3^2 - m(w+k_1)=0\,.
\end{equation}
The linear terms indicate that these expressions cannot be represented by
pseudo-Riemannian metrics. 
\begin{figure*}[h]
$\!\!\!\!$\begin{tabular}{ccc}
\begin{minipage}{3.0in}
\centering
\includegraphics[angle=0,width=3.0in]{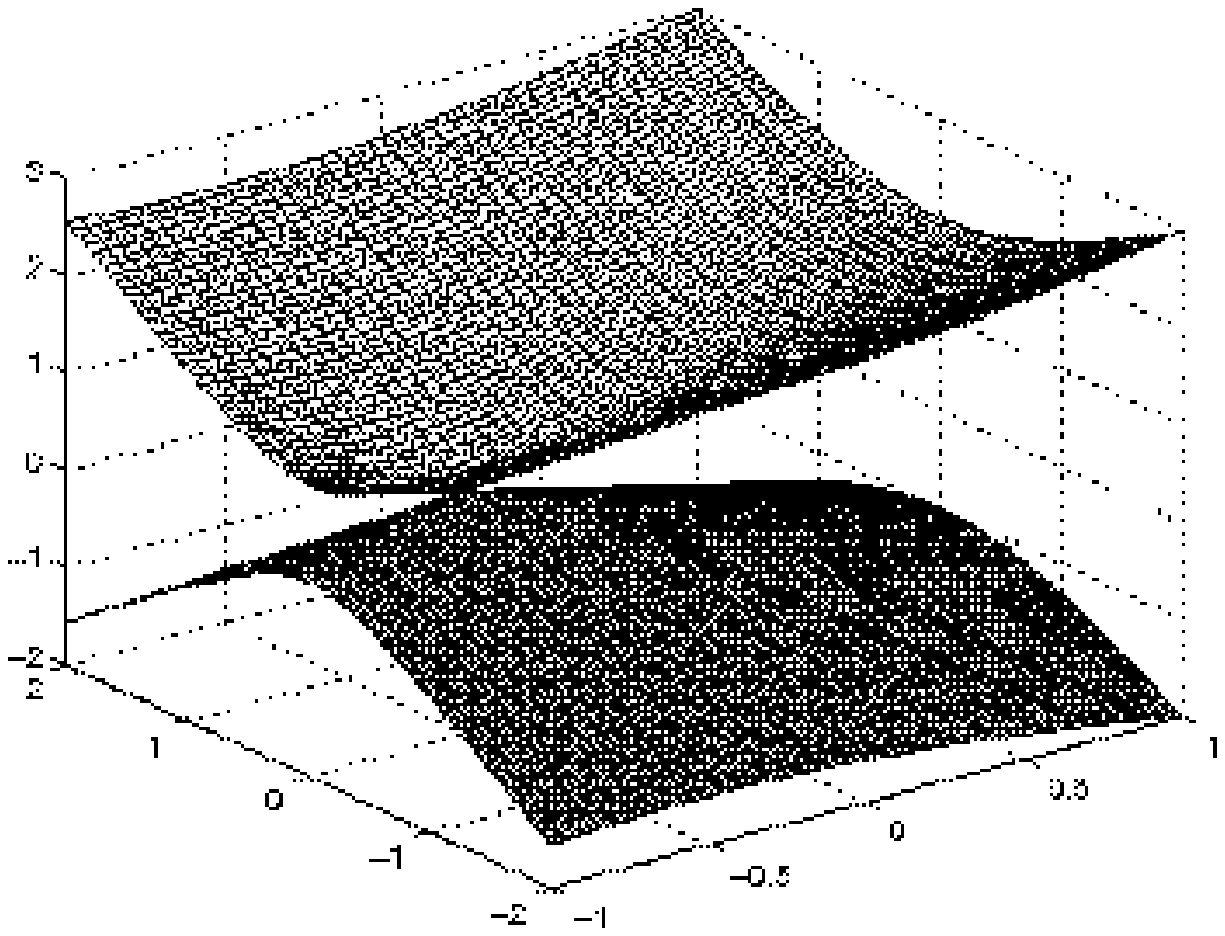}
\end{minipage}
&
\begin{minipage}{3.0in}
\centering
\includegraphics[angle=0,width=3.0in]{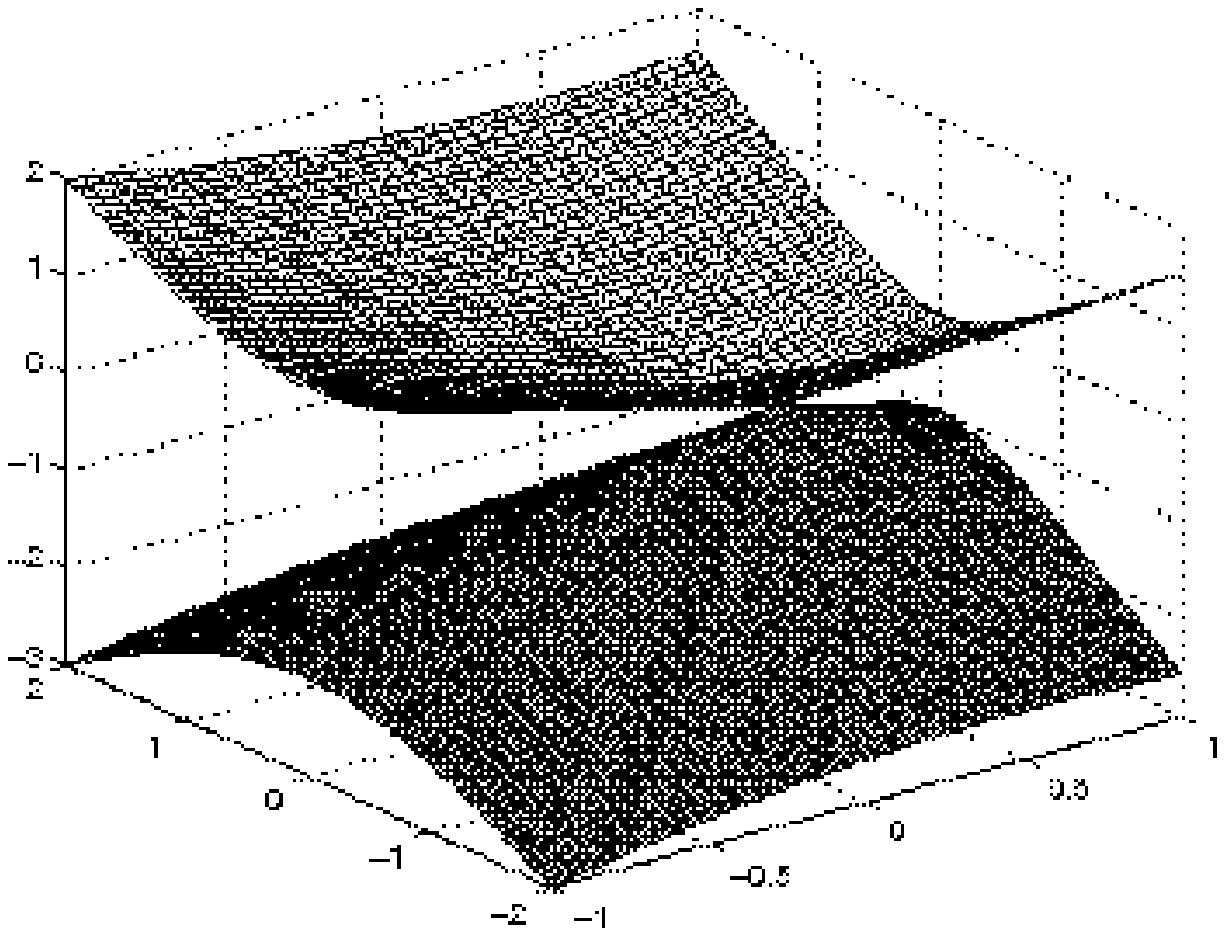}
\end{minipage}
\end{tabular}
\caption{The graphs represent two light hypersurfaces of the lightlike axion
modified electrodynamics } \label{cone3}
\end{figure*}

The equations (\ref{disp-nul3}) have four   solutions
  \begin{equation}\label{sol-quad1}
{}^{(1)}w=\frac m2+ \sqrt{\frac{m^2}4+k^2+mk\cos \alpha}
\qquad {}^{(2)}w=\frac m2- \sqrt{\frac{m^2}4+k^2+mk\cos \alpha}\,,
 \end{equation}
 and
  \begin{equation}\label{sol-quad2}
{}^{(3)}w=-\frac m2+ \sqrt{\frac{m^2}4+k^2-mk\cos \alpha}\qquad
{}^{(4)}w=-\frac m2- \sqrt{\frac{m^2}4+k^2-mk\cos \alpha}\,.
 \end{equation}
 These expressions  are real for every values of the parameters.
 Consequently,  for the axion field with lightlike derivative,
  the runaway solutions are absent.

  The light hypersurfaces are depicted on Fig.3. The first picture corresponds to (\ref{hyper-nul3}), while the second one is for (\ref{hyper-nul3x}).  The future and the past surfaces are contacted at points
  \begin{equation}\label{spheres}
 w=\pm\frac m2\,,\quad k_1=\pm\frac m2\,,\quad k_2=0\,,\quad k_3=0\,.
  \end{equation}
 This topological structure is also different from the standard light cone structure.

The  electromagnetic waves propagate with two different
 phase velocities
\begin{equation}\label{phase-vel1}
{}^{(1)}v_p=\frac m{2k}+\sqrt{\frac {m^2}{4k^2}+\frac m{k}\cos\alpha+1}\,,
\end{equation}
and
\begin{equation}\label{phase-vel2}
{}^{(2)}v_p=-\frac m{2k}+\sqrt{\frac {m^2}{4k^2}-\frac m{k}\cos\alpha+1}\,.
\end{equation}
When the dimensionless parameter $m/k$ increases, i.e., for small $k$,
 the first velocity (\ref{phase-vel1}) goes to infinity, while the second one
 (\ref{phase-vel2}) tends to  zero.

The group velocities are expressed as
\begin{equation}\label{group-vel1}
{}^{(1)}v_g=\frac {m\cos\alpha+2k} {\sqrt{ {m^2}+4km\cos\alpha+ 4k^2}}\,,
\end{equation}
and
\begin{equation}\label{group-vel2}
{}^{(2)}v_g=\frac {m\cos\alpha- 2k} {\sqrt{ {m^2}+4km\cos\alpha- 4k^2}}\,.
\end{equation}
For small $k$ they tend to the same value $\cos\alpha\le 1$.
\section{Conclusions}
We have considered a general phenomenological model of axion modified electrodynamics.
It is shown that the  axion modified electrodynamics can be treated as
 a special case of premetric electrodynamics.
 In this formalism, the axion field is not involved as an additional
 Chern-Simon term. Alternatively, it emerges as
 an irreducible part of a general constitutive tensor.
 We have derived a covariant dispersion relation of axion electrodynamics.
 For a varying axion field, it yields a modification of light cone.
 The birefringence effect indicates violation of the Lorentz invariance for
 timelike, spacelike and null covector of axion field derivatives.
This effect, however, completely different from the ordinary birefringence
appearing in  classical optics and in the premetric electrodynamics.
It can be explicitly seen from the fact that the topological structure
of the light hypersurface is different from the ordinary light cone structure.
In addition, the optical metrics are not homogeneous in the wave covector so they are even non-Finslerian. 

\begin{acknowledgements}
My thanks to Roman Jackiw, Friedrich Hehl and Volker Perlick for useful discussion. My deep acknowledgments to Yuri Obukhov for valuable corrections.
\end{acknowledgements}

\begin{appendix}
\section*{Appendix A. Tensors and pseudotensors}
Since the axion field possesses a special transformational behavior, it is useful to have  a precise meaning of all quantities involved in the formalism. Although the notion of weighted tensors and pseudotensors is a classical subject, see \cite{Sch} and also \cite{Birkbook} for a modern treatment, some confusions in the basic definitions can be found in literature.  We will characterize  the  geometrical quantities accordingly  to the transformation properties of their components relative to  transformations of coordinates.

Let be given a smooth transformation
\begin{equation}\label{trans}
x^i\to x^{i'}=f^{i'}(x^i)
\end{equation}
 with a transformation matrix
\begin{equation}\label{trans-mat}
L^{i'}{}_i=\frac{\partial f^{i'}(x^i)}{\partial x^{i}}
\end{equation}
 and its inverse $L_{i'}{}^i$.
Denote the determinant of the transformation matrix by $J$. It is expressed by
\begin{equation}\label{det-trans-mat}
J={\rm det}(L^{i'}{}_i)=\frac 1{4!} \epsilon_{i'j'k'l'}\epsilon^{ijkl}
L^{i'}{}_i \,L^{j'}{}_j\,L^{k'}{}_k\,L^{l'}{}_l\,,
\end{equation}
where $\epsilon^{ijkl}$ and  $\epsilon_{ijkl}$ are the  permutation symbols.  They take the constant values $0, \pm 1$ in all coordinate systems.
The relation (\ref{det-trans-mat}) is equivalent to
\begin{equation}\label{det-trans-mat-x}
\epsilon^{i'j'k'l'}J=\epsilon^{ijkl}
L^{i'}{}_i\,L^{j'}{}_j\,L^{k'}{}_k\,L^{l'}{}_l\,,
\end{equation}
and
\begin{equation}\label{det-trans-mat-xx}
\epsilon_{i'j'k'l'}=J\epsilon_{ijkl}
L_{i'}{}^i\,L_{j'}{}^j\,L_{k'}{}^k\,L_{l'}{}^l\,.
\end{equation}
Recall, see \cite{Sch}, the definitions of the extended tensorial objects. 

\paragraph{Ordinary tensor}
has a set of components which  are transformed as
\begin{equation}\label{tensor}
T^{i'\cdots}{}_{j'\cdots} \to L^{i'}{}_{i}\cdots L_{j'}{}^{j}\cdots T^{i\cdots}{}_{j\cdots} \,.
\end{equation}
\paragraph{Tensor density  of weight $k$ }
is a set of components which  are transformed as
\begin{equation}\label{tensor-dens}
T^{i'\cdots}{}_{j'\cdots} \to \frac1{J^k}L^{i'}{}_{i}\cdots L_{j'}{}^{j}\cdots T^{i\cdots}{}_{j\cdots} \,.
\end{equation}
So the ordinary tensors are tensor densities of zero weight.
\paragraph{Pseudotensor}
has a set of components which  are transformed as
\begin{equation}\label{pstensor}
T^{i'\cdots}{}_{j'\cdots} \to ({\rm sgn }\,J) L^{i'}{}_{i}\cdots L_{j'}{}^{j}\cdots T^{i\cdots}{}_{j\cdots} \,.
\end{equation}

\paragraph{Pseudotensor density  of weight $k$ }
is a set of components which  are transformed as
\begin{equation}\label{pstensor-dens}
T^{i'\cdots}{}_{j'\cdots} \to  \frac{{\rm sgn }\,J} {J^k}\, L^{i'}{}_{i}\cdots L_{j}{}^{j'}\cdots T^{i\cdots}{}_{j\cdots} \,.
\end{equation}
The ordinary pseudotensors are of zero weight.

Let us start with an action
\begin{equation}\label{act}
{\mathcal A}=\int_M L\,vol\,,
\end{equation}
which is an ordinary real number.
Its integrand, $L\,vol$, has to be a scalar valued invariant volume element, i.e., a pseudoscalar density. In the formalism of differential forms, it is an odd (twisted) 4-form,  see for instance \cite{Gronwald:1997ei}. Here $L$ is a scalar valued function (Lagrangian) while $vol$ is a special invariant volume element defined from the geometric quantities. On a $4D$ pseudo-Riemannian manifold, the standard invariant volume element is defined by the determinant of the metric tensor $g={\rm det}(g_{ij})$
\begin{equation}\label{Ri-vol}
^{(Riem)}vol=\sqrt{-g}d^4x\,,\qquad d^4x=dx^0dx^1dx^2dx^3\,.
\end{equation}
In a pure tensorial form, it is equivalently rewritten as
\begin{equation}\label{Ri-vol1}
 ^{(Riem)}vol=\sqrt{-g}\left(\frac1{4!} \epsilon_{ijkl}dx^idx^jdx^kdx^l\right)\,.
\end{equation}

Due to (\ref{det-trans-mat-x}), the permutation symbol $\epsilon^{ijkl}$ is a tensor density of weight $(+1)$. Its "inverse" $\epsilon_{ijkl}$ is a tensor density of weight $(-1)$.
Consequently $d^4x$ is a scalar density of weight $(-1)$.

From the ordinary transformation law for the metric tensor
 \begin{equation}\label{metr}
 g_{i'j'}=g_{ij}L^i{}_{i'}L^j{}_{j'}\,,
 \end{equation}
 one readily has the transformation law for the determinant and its square root
 \begin{equation}\label{metr-det}
 g'=\frac 1{J^2}g\,, \qquad \sqrt{-g'}=\frac1{|J|}\sqrt{-g}=\frac{sgn(J)}{J}\sqrt{-g}\,.
 \end{equation}
 Thus $g$ is a scalar density of weight $(+2)$ while $\sqrt{-g}$ is a pseudoscalar density of weight $(+1)$.

 One builds from these two objects a quantity $\tilde{\epsilon}_{ijkl}=\sqrt{-g}\epsilon_{ijkl}$, which is a (non-weighted)  pseudotensor.  Observe that the integrand of the action $L\, vol$ has a proper transformation behavior, it is a pseudoscalar of zero weight.   Since the expression $L\sqrt{-g}$ is the subject of variation, this pseudoscalar density of weight $(+1)$ is often referred to as the Lagrangian density.

Let us return now to the premetric electrodynamics action
\begin{equation}\label{lagr1x}
{\mathcal A}=\int_M \left(F_{ij}{\mathcal H}^{ij} + A_i{\mathcal J}^{i}\right)d^4x\,,
\end{equation}
and, equivalently,
\begin{equation}\label{lagr2x}
{\mathcal A}=\int_M \left(\frac 12 \chi^{ijkl}F_{ij}F_{kl}+ A_i{\mathcal J}^i\right)d^4x\,.
\end{equation}
Since $d^4x$ is a scalar density of weight $(-1)$, the scalar integrand (the expression in the parenthesis) has to be treated as a pseudoscalar density of  weight $(+1)$.
It is constructed from the  ordinary tensors $F_{ij}$ and $A_i$ and the
pseudotensor densities of  weight $(+1)$ --- ${\mathcal H}^{ij}$ and ${\mathcal J}^{i}$.
On a pseudo-Euclidean manifold, the ordinary tensors $H^{ij}$ and $J^i$ are extracted from them by ${\mathcal H}^{ij}=H^{ij}\sqrt{-g}$ and
${\mathcal J}^{i}=J^i\sqrt{-g}$.
Consequently, $\chi^{ijkl}$,  which is defined by ${\mathcal H}^{ij}=\chi^{ijkl}F_{kl}$,  is a pseudotensor density of  weight $(+1)$.

In axion electrodynamics, the constitutive tensor is given by
\begin{equation}\label{axion}
^{\tt (Max)}\chi^{ijkl}=\left(g^{ik}g^{jl}-g^{il}g^{jk}\right)\sqrt{-g}+ \psi\epsilon^{ijkl}\,.
\end{equation}
Since $\sqrt{-g}$ s a pseudoscalar density of weight $(+1)$, the first term is a pseudotensor density  of weight $(+1)$. In the second term, $\epsilon^{ijkl}$ is a tensor density of weight $(+1)$. Consequently, $\psi$ is a  pseudoscalar. In physics literature, it is called axion.

The matrix $M^{ij}=\chi^{ijkl}q_kq_l$ is a pseudotensor density of weight $(+1)$. Its determinant is a scalar density of weight $(+2)$. Consequently the adjoint matrix $A_{ij}$ is a pseudotensor density of weight $(+1)$ while  the function $\lambda(q)$ is a pseudoscalar density of weight $(+1)$. The photon propagator derived for premetric electrodynamics in
\cite{Itin:2007av} is a pseudotensor of weight $(-1)$.
\end{appendix}

\end{document}